\documentclass[twocolumn,showpacs,prb,amsmath,amssymb,aps,floatfix]{revtex4-1}

\usepackage{graphicx,subfigure}
\usepackage{dcolumn}
\usepackage{bm}
\usepackage{gensymb}
\usepackage{wasysym}

\begin{document}

\title{A model for the epitaxial growth of graphene on 6H-SiC}

\author{Fan Ming}
\author{Andrew Zangwill}%
 \email{andrew.zangwill@physics.gatech.edu}
\affiliation{%
School of Physics, Georgia Institute of Technology, Atlanta, GA 30332, USA
}%

\date{\today}

\begin{abstract}
We introduce a kinetic model for the growth of epitaxial graphene on 6H-SiC. The model applies to vicinal surfaces composed of half-unit-cell height steps where experiment shows that step flow sublimation of SiC promotes the formation and growth of graphene strips parallel to the step edges. The model parameters are effective energy barriers for the nucleation and subsequent propagation of graphene at the step edges.  Using both rate equations and kinetic Monte Carlo simulations, two  distinct growth regimes emerge from a study of the layer coverage and distribution of top-layer graphene strip widths as a function of total coverage, vicinal angle, and the model parameters.  One regime is dominated by the coalescence of strips. The other regime is dominated by a novel ``climb-over'' process which facilitates the propagation of graphene from one terrace to the next. Comparing our results to scanning microscopy studies will provide the first quantitative insights into the kinetics of growth for this unique epitaxial system.
\end{abstract}


\maketitle

\section{Introduction}
Of the various ways known to produce graphene, it is increasingly likely that only  graphene grown epitaxially on silicon carbide will play an important role in post-CMOS microelectronics \cite{First2010,Sprinkle2010}. This system is unique among epitaxial growth systems because there is no deposition flux. Instead, silicon atoms sublime from SiC(0001) and SiC(000$\bar{\rm 1}$) at high temperature and the carbon atoms left behind recrystallize into graphene. Microscopy studies of vicinal 6H-SiC surfaces show that graphene nucleates at step edges and that the subsequent morphology depends strongly on growth conditions, vicinality, and whether the step heights are equal to one, two, or three Si-C bilayers \cite{Lauffer2008,Virojanadara2008,Hupalo2009,Emtsev2009,Bolen2009,Ohta2010}. For example, growth from single-bilayer steps produces a complex, finger-like graphene morphology \cite{Hupalo2009,Ohta2010} while step-flow growth from triple-bilayer (half-unit-cell) height steps produces long, straight strips parallel to the steps whose widths increase as growth proceeds \cite{Lauffer2008,Ohta2010}. The difference comes from the fact that three bilayers of SiC must desorb to liberate a sufficient number of carbons atoms to cover the sublimed area with one layer of graphene.

The calculations reported in this article aim to (i) provide experimenters with a simple and convenient way to characterize the changes they see in surface morphology when growth conditions change; (ii) identify a statistical measure of sub-monolayer growth which identifies whether graphene step-flow growth is limited by nucleation at steps or by propagation on terraces; and (iii) provide physical insight into the competition between graphene strip coalescence and a new kinetic process (unique to this system) which we call ``climb-over''. Our principal theoretical tool is a phenomenological kinetic Monte Carlo (KMC) model of the sort used to study the growth kinetics of III-V semiconductors \cite{Joyce1994}. When coupled closely with experiment, this approach produced a decade of valuable insights before simulations based on total energy calculations of energy barriers for III-V systems became possible \cite{Kratzer2002}. For our problem, first-principles KMC is impossible because the structure of the ``buffer layer'' at the graphene/SiC interface is controversial and the structure of steps on this buffer layer is unknown \cite{Hass2008}. On the other hand, the fact that graphene grows in strips from triple bilayer steps means that a one-dimensional model is a good first approximation and the distribution of these strip widths will play a prominent role in what follows. A mean-field rate-equation analysis in the Appendix provides further insight into our proposed model and the KMC results.

\section{Kinetic Monte Carlo Model}
\begin{figure}
\includegraphics[scale=0.38]{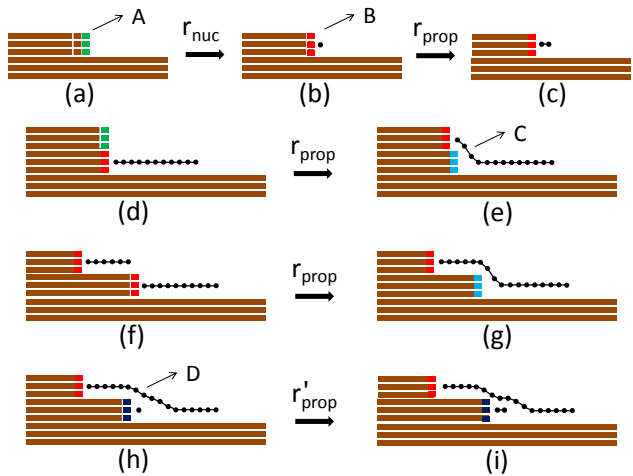}
\caption{Kinetic processes allowed in the KMC simulation. ThMK6DTY-T7S9SCe steps marked A (green), B (red), C (blue), and D (purple) play a role in the rate theory reported in the Appendix.}
\label{model}
\end{figure}

Fig.~\ref{model} shows the various processes we consider for a vicinal surface of 6H-SiC composed exclusively of half-unit-cell height steps. Each process involves the replacement of a unit area of SiC triple bilayer by a unit area of graphene. Consistent with the coarse-grained nature of the our model, we do not concern ourselves with atomistic details and simply assume that all exposed SiC terraces spontaneously reconstruct to the carbon-rich ``buffer layer'' known to form on both 6H-SiC($0001$) and 6H-SiC($000\bar{1}$) \cite{Hass2008}. Fig.~\ref{model}(a)-(b) shows the nucleation of graphene at a SiC step with no graphene nearest neighbors. This occurs in our model at a rate $r_{\rm nuc} = \nu_0 \exp(-E_{\rm nuc}/kT)$, where $\nu_0 \approx 10^{12} s^{-1}$ is an attempt frequency and $T$ is the substrate temperature. Two points are worth noting. First, $E_{\rm nuc}$ is an {\it effective} energy parameter which accounts for the combined effects of Si atoms sublimation, C atoms re-crystallization, and graphene growth along the step edge. Second, a variation of our model could allow additional SiC to sublime before a stable graphene nucleus forms. This influences the predicted distribution of strip widths and, like the corresponding problem of critical island sizes in conventional epitaxial growth, comparison with experiment provides microscopic information that is nearly impossible to learn any other way \cite{Evans2006}.

After nucleation, graphene growth continues by dissolution of the adjacent SiC step at a rate  $r_{\rm prop} = \nu_0 \exp(-E_{\rm prop}/kT)$ [Fig.~\ref{model}(b)-(c)]. Propagation occurs only at SiC steps that are bounded by a graphene strip. Two fates are possible for such a strip. One is that the propagating strip runs into another SiC step and creates a step bunch of two triple bilayer steps [Fig.~\ref{model}(d))]. If this happens, the strip can ``climb over" the upper terrace at the rate $r_{\rm prop}$ [Fig.~\ref{model}(e)]. Another possibility is that the propagating strip meets another strip on the upper terrace [Fig.~\ref{model}(f)]. In this case, our KMC simulation coalesces the two strips at the rate $r_{\rm prop}$ [Fig.~\ref{model}(g)]. Nucleation of a covered graphene layer at a covered SiC step occurs at the rate $r_{\rm nuc}$ [Fig.~\ref{model}(g)-(h)]. Propagation of a covered graphene layer occurs at the rate $r_{\rm prop}$ or (for some of the simulations reported below) at the slower rate $r'_{\rm prop}$ [Fig.~\ref{model}(i)]. The later growth continues in the same way as the first graphene layer.

We use a standard KMC algorithm \cite{Voter2007} to simulate growth on vicinal SiC surfaces composed of (typically) 5000 steps with periodic boundary conditions. At least 100 independent runs were averaged to obtain statistically significant results. The vicinal angle $\phi=\tan^{-1}(3/W)$, where $W$ is the terrace width. We begin our discussion with $\Theta_i$, the graphene coverage of layer $i$, as a function of the total graphene coverage $\Theta = \sum_{i} {i\Theta_i}$. These quantities are accessible to spatial-averaging experimental probes and our model energy parameters should provide a simple and convenient way for experimenters to characterize variations in observed morphology with growth conditions. Later, we will turn to the distribution of graphene strip widths as a quantity which scanning microscopy can exploit to learn the relative importance of competing surface kinetic processes during growth.

\section{Layer Coverage and Growth Time}
\begin{figure}
\includegraphics[scale=0.35]{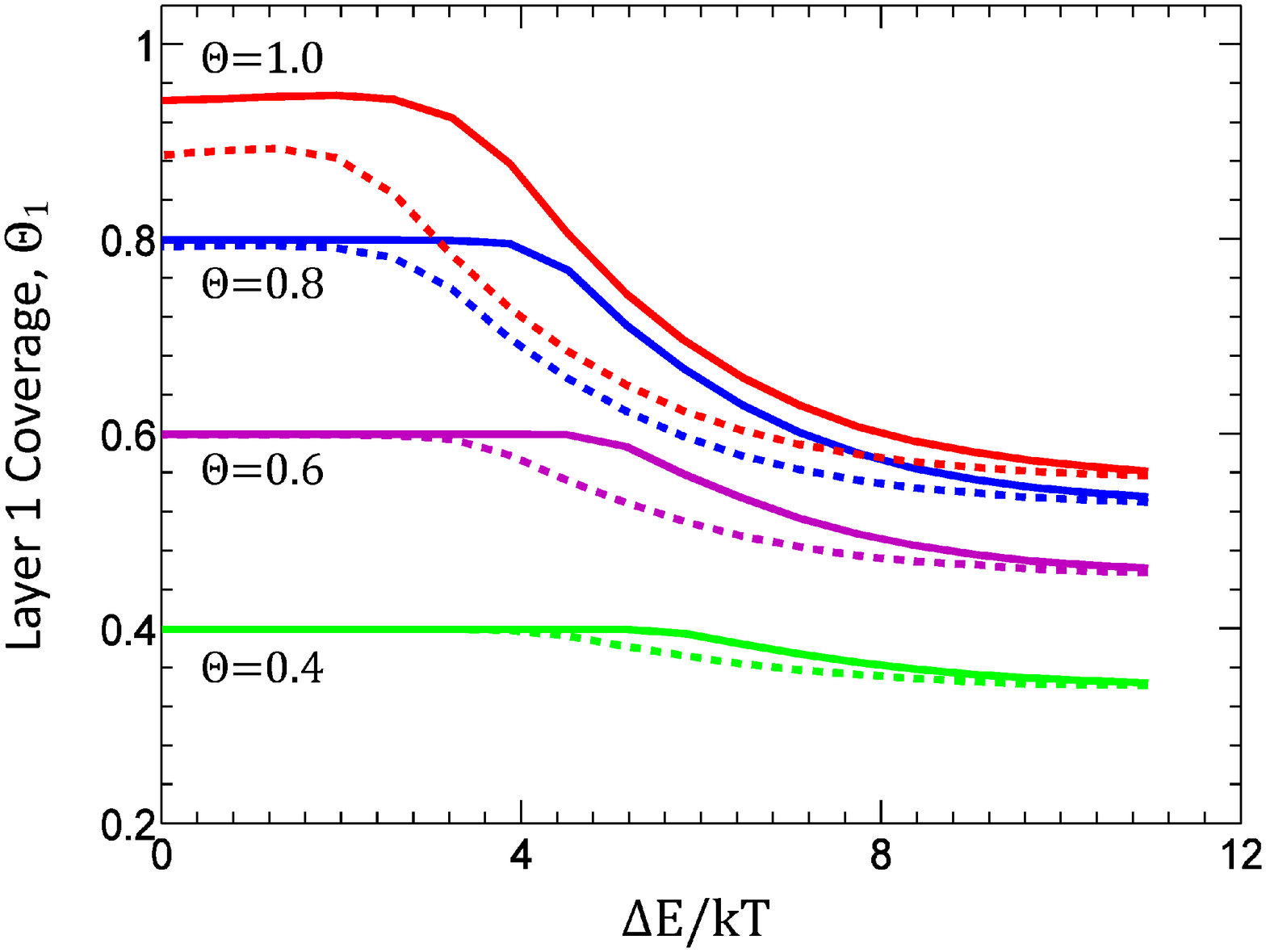}
\caption{Layer 1 coverage as a function of the energy barrier difference $\Delta E$ and vicinality $\phi$. Solid and dashed lines correspond to $\phi = 0.9^{\circ}$ and $\phi = 3.4^{\circ}$, respectively.}
\label{layer1-layer2}
\end{figure}

Fig.~\ref{layer1-layer2} shows simulation results for $\Theta_1$ as a function of $\Delta E = E_{\rm nuc}-E_{\rm prop}$ with $r'_{\rm prop}=r_{\rm prop}$ for four different values of total coverage $\Theta$ and two choices for the vicinal angle $\phi$. The rather counter-intuitive behavior that $\Theta_1$ {\it decreases} as $\Delta E$ {\it increases} for fixed $\Theta$ can be understood as follows. When $\Delta E$ is large, propagation of existing graphene strips  is relatively more likely than the nucleation of new graphene strips, and fewer graphene strips can form. Many strips undergo the ``climb over" process [Fig.~\ref{model}(d)-(e)] when the width of the graphene strips passes the terrace width, thereby creating many nucleation sites for second layer growth. The net result is that nucleation of second layer graphene begins earlier. Because these strips grow for a longer time, the total second layer coverage is larger. Fig.~\ref{layer1-layer2} also implies that better surface homogeneity can be achieved by increasing the substrate temperature and decreasing the substrate miscut angle. This conclusion is consistent with the observations reported in Ref.~\onlinecite{Virojanadara2010}. When $\Delta E$ is further increased, the number of nucleation events for both the first layer and second layer are greatly reduced, and eventually the competition between the two layers is balanced. This effect produces the lower plateaus in Fig.~\ref{layer1-layer2}. A rate equation analysis described in the Appendix  provides another way to understand this behavior of our model.

\begin{figure}
\includegraphics[scale=0.35]{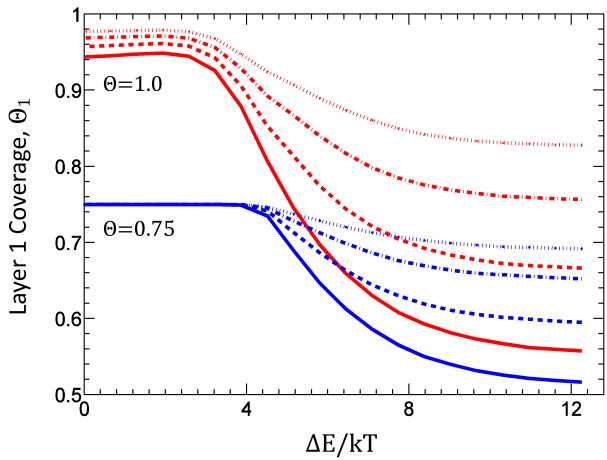}
\caption{Layer 1 coverage as a function of $\Delta E$ for different values of $\Delta E'$. $\Delta E'/kT = 0, 0.6, 1.3, 1.9$ applies to the solid curves, dashed curves, dashed-dotted curves, and dotted curves, respectively. The vicinal angle $\phi = 0.9^{\circ}$.}
\label{layer2barrier}
\end{figure}

In principle, experimental data for $\Theta$ and $\Theta_1$ can be compared with the curves in Fig.~\ref{layer1-layer2} (outside the plateau regime) to extract a value for $\Delta E$. However, because it is surely harder for Si atoms to escape from SiC when they are covered by a graphene layer than when they are not, we introduce a second layer propagation barrier $E'_{\rm prop}>E_{\rm prop}$. The corresponding rate for second layer propagation is $r'_{\rm prop} = \nu_0 \exp(-E'_{\rm prop}/kT)$ and we define $\Delta E' = E'_{\rm prop} - E_{\rm prop}$. We retain the equality of the first and second layer nucleation barriers for simplicity\footnote{The authors thank Prof. Thomas Seyller of the University of Erlangen for suggesting this refinement. See also Ref.~\onlinecite{Hupalo2009}.}. We also forbid the growth of layer 3. Fig.~\ref{layer2barrier} shows that as $\Delta E'$ increases, the first layer coverage increases substantially, as might be expected. We now have a three parameter problem and experimental data for $\Theta_1$ at two values of total coverage $\Theta$ can be used to extract values for  $\Delta E$ and $\Delta E'$ from Fig.~\ref{layer2barrier}.

\begin{figure}
\includegraphics[scale=0.35]{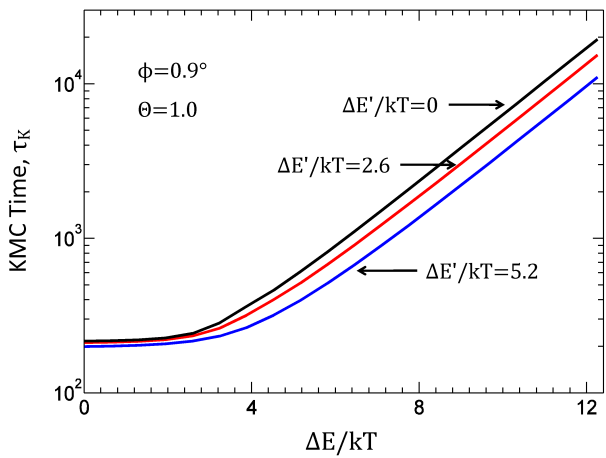}
\caption{KMC time $\tau_K$ as a function of the energy barrier difference $\Delta E$ and $\Delta E'$ at a fixed total coverage.}
\label{total}
\end{figure}

It remains to deduce values of  $E_{\rm nuc}$, $E_{\rm prop}$ and $E'_{\rm prop}$ individually. This can be done using the experimental growth time for a given total coverage because $\tau_K = t_E r_{\rm prop}$ relates the dimensionless KMC simulation time to the experimental growth time $t_E$. Fig.~\ref{total} shows $\tau_K$ as a function of $\Delta E$ for different values of $\Delta E'$. This graph (or a similar one obtained for a different choice of $\Theta$ and $\phi$) permits $E_{\rm prop}$ to be extracted from the values of $\Delta E$ and $\Delta E'$ determined earlier from the layer coverage curves. The two other energy parameters follow immediately.

\section{Strip Width Distribution}
We turn now to the distribution of first-layer graphene strip widths. This statistical quantity probes more deeply into the competition between nucleation and propagation and between coalescence and climb-over. It also provides another way to extract $\Delta E$ and to understand the cross-over from the low-$\Delta E$ plateau to the high-$\Delta E$ plateau in Fig.~\ref{layer1-layer2}. Compared to the coverage curves, this distribution is much more sensitive to $\Delta E$ and much less sensitive to $\Delta E'$. For that reason, we set the latter equal to zero in what follows.

\begin{figure}
\includegraphics[scale=0.56]{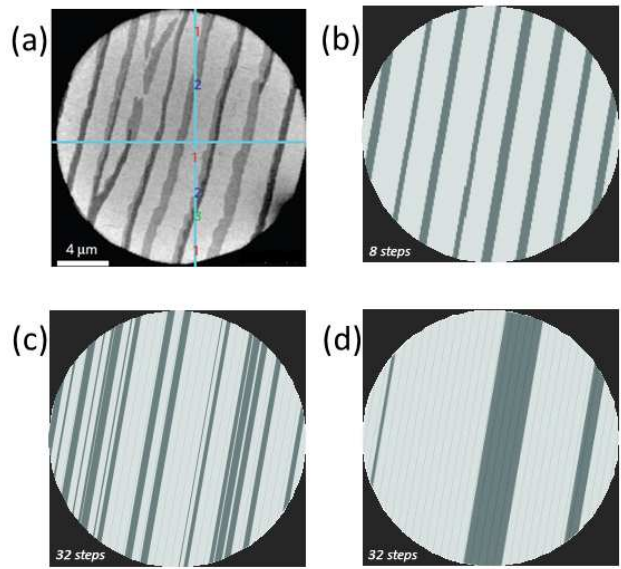}
\caption{(a) LEEM image of graphene grown on vicinal 6H-SiC($0001$) from Ref.~\onlinecite{Emtsev2009}. Regions covered by one, two, and three layers of graphene are shown as light, moderate, and dark gray, respectively. The latter two occur at SiC step edges. (b)-(d), KMC simulation images of monolayer graphene strips with $\Delta E/kT = 0$, $5.8$, and $11.6$, respectively. The total coverage $\Theta = 0.25$. Light grey lines and the right edges of graphene stripes are SiC steps. The vicinal angle $\phi=0.9^\circ$.}
\label{image}
\end{figure}

\begin{figure}
\includegraphics[scale=0.9]{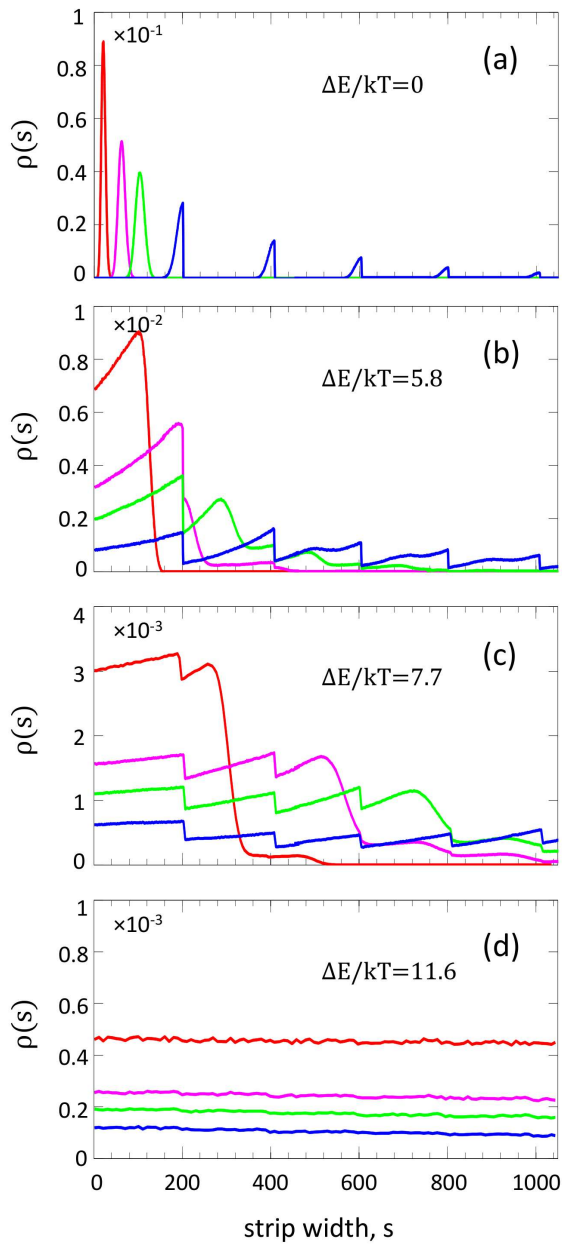}
\caption{Graphene strip width distribution $\rho(s)$ for different $\Delta E$ and total coverages with $\phi=0.9^\circ$. Different color lines correspond to $\Theta=0.1$ (red), $0.3$ (magenta), $0.5$ (green) and $1.0$ (blue), respectively. The terrace width is $W=200$.}
\label{distr}
\end{figure}

Fig.~\ref{image}(a) shows a LEEM image \cite{Emtsev2009} where the terraces are mostly covered by a single monolayer of graphene (light gray). Very near the step edges, strips composed of two (moderate gray) and three (dark gray) layers of graphene are apparent. Fig.~\ref{image}(b) shows a KMC simulated morphology (with $\Delta E=0$) which looks quite similar to Fig.~\ref{image}(a). The graphene strip morphology changes significantly as $\Delta E$ increases in  Fig.~\ref{image}(c) and (d): the  number of graphene strips decreases and many of them cover many SiC steps. Note also the change in scale from Fig.~\ref{image}(a). To quantify this morphological change, Fig.~\ref{distr} plots $\rho(s)$, the normalized distribution of strips with width $s$, for different choices of $\Delta E$ and $\Theta$. The terrace width here is $W=200$. When $\Delta E/kT = 0$ [Figure~\ref{distr}(a)], the distribution is Poisson because graphene nucleates at almost all the SiC steps simultaneously. The mean strip width is $W\Theta$ in the interval $[0, W]$. However, when the coverage $\Theta>0.8$ (blue line in Fig.~\ref{distr}(a)), the leading edge of the Poisson distribution begins to cross the terrace width $W$, a few strips disappear by the ``coalescence" mechanism [Fig.~\ref{model}(f)-(g)], and a few strips with widths close to $2W$ form. As a result, the strip width distribution is abruptly cut-off at the terrace width and the Poisson distribution repeats (with a much decreased peak amplitude) in the width interval $[W, 2W]$. The distribution moves farther across the terrace width boundary when $\Theta$ increases further.

The general behavior of $\rho(s)$ with increasing coverage persists when the value of $\Delta E$ increases. However, a larger value of $\Delta E$ implies that some graphene strips nucleate earlier than others. This leads to a shift to the right in the peak position seen in Fig.~\ref{distr}(a)-(c) for the same coverage. The increasing time delay between consecutive nucleation events similarly produces a distinct broadening of the distribution curves. Eventually, for large enough $\Delta E$, the distribution curves become uniform [Fig.~\ref{distr}(d)] in the scale we consider. This occurs when the graphene strips propagate so rapidly (relative to nucleation) that the step edges are no longer distinguishable.

Finally, we return to the vicinal angle dependence of the coverage curves plotted in Fig.~\ref{layer1-layer2}. The $\rho(s)$ results above imply that the transition between the two horizontal plateaus in these graphs as $\Delta E$ increases reflects a transition from a Poisson distribution to a uniform distribution of graphene strip widths. In the Poisson regime, the terrace width only affects the coverage distribution at late times when the maximum graphene strip width passes $W$. Therefore, a change in the vicinal angle $\phi$ only changes the coverage distribution for large $\Theta$. This may be contrasted with the uniform regime, where the graphene strips grow so rapidly that they are not hindered by the SiC step edges. In this case, the coverage distribution does not depend on $\phi$ at all. Nevertheless, as we see from Fig.~\ref{layer1-layer2}, for fixed $\Theta$, $\Theta_1$ tend to be larger for smaller vicinal angle $\phi$. This is so because the standard deviation divided by the terrace width for a Poisson distribution is $\sqrt{W\Theta}/W \sim 1/\sqrt{W}$, which implies that a smaller vicinal angle leads to a relatively narrower distribution of strip widths. In the limit when all the graphene strips are about the same width, coalescence events occur only very near $\Theta=1$ and there is essentially no second layer growth. This supports our previous statement that better surface uniformity can be achieved by using a more singular surface.

\section{CONCLUSION}
In summary, we have developed a one-dimensional kinetic Monte Carlo model to study the epitaxial growth of graphene by the step flow sublimation of vicinal surfaces of 6H-SiC(0001). The layer coverages and the distribution of graphene strip widths were found to depend more or less strongly on the relative sizes of the effective energy barriers for graphene nucleation, first layer propagation, and second layer propagation. The cross over of the distribution from Poisson to uniform as the nucleation barrier increases clearly shows that there are two distinctive growth regimes, one dominated by  ``coalescence" processes and one dominated by  ``climb over" processes. The ``climb over" processes have the effect of increasing the graphene surface inhomogeneity. It will be interesting to compare these simulation results with experimental measurements to see how the effective energy barriers depend on growth parameters like the partial pressure of silicon in the growth chamber. Future simulations studies will examine the ``kinetic roughening'' of this new model of epitaxial growth \cite{Michely2004}, and the ability of the model to rationalize the non-uniform layer thicknesses observed when graphene grows on spontaneously facetted SiC substrates \cite{Norimatsu2010}.

\section{ACKNOWLEDGMENTS}
The authors thank Ed Conrad, Walt de Heer, Phil First, and Dimitri Vvedensky for their insights. This research was supported in part by the National Science Foundation through TeraGrid resources provided by Texas Advanced Computing Center (TACC) under Grant No. TG-PHY100006. Fan Ming was supported by the MRSEC program of the National Science Foundation under Grant No. DMR-0820382.

\section{APPENDIX: RATE EQUATION ANALYSIS}
This Appendix presents a mean-field rate-equation analysis to provide further understanding of our graphene growth model.  The color coding in Fig.~\ref{model} identifies four basic types of steps: (A) a bare SiC step; (B) a step connected to a layer 1 graphene segment; (C) a step that is carpeted by a continuous layer of graphene; and (D) a step that is connected to a layer 2 graphene segment. We let  $n_A$, $n_B$, $n_C$, and $n_D$ be the number of steps of each type, so $n=n_A + n_B + n_C + n_D$ is the total number of steps and $L=nW$ is the system size. Then, if $p_d n_B$ is the number of B-steps with an A-step immediately above [Figure~\ref{model}(d)] and $p_f n_B$ is the number of B-steps with a graphene segment immediately above [Figure~\ref{model}(f)], an approximate description of the epitaxial graphene growth processes is
\begin{eqnarray}
\frac{d n_A}{d \Theta} &=& -r_1 n_A - r_2 p_d n_B  \label{nA} \\
\frac{d n_B}{d \Theta} &=& r_1 n_A -  r_2 p_f n_B  \label{nB} \\
\frac{d n_C}{d \Theta} &=& - r_1 n_C + r_2 (p_d+p_f)n_B  \label{nC} \\
\frac{d n_D}{d \Theta} &=& r_1 n_C  \label{nD}
\end{eqnarray}
where $r_1=r_{\rm nuc}L/r_{tot}$, $r_2=r_{\rm prop} L/r_{tot}$, and $r_{tot} = r_{\rm nuc} n_A + r_{\rm prop} n_B + r_{\rm nuc} n_C + r'_{\rm prop} n_D$.

Eq.~(\ref{nA}) says that A-steps (green) are lost by first-layer nucleation events and by climb-over events. Eq.~(\ref{nB}) says that B-steps (red) are created by nucleation events and lost by coalescence events. Eq.~(\ref{nC}) says that C-steps (blue) are lost by second-layer nucleation events and created by both climb-over and coalescence events. Eq.~(\ref{nD}) says that D-steps (purple) are created by second-layer nucleation events. We note that a climb-over event does not change the number of B-steps.

We consider two limits where $p_d$ and $p_f$ can be estimated. The first limit is  $r_{\rm nuc} \ll r_{\rm prop}$ where first-layer nucleation events are rare. Climb-over is frequent and coalescence infrequent. These conditions imply, in turn, that  $p_f \ll 1$ and $p_d \approx 1/W$. The second of these is true because, when the coverage is fixed and the propagation rate is very fast, the length (modulo $W$) of the graphene segment connected to a B-step takes every value between one and the terrace length $W$. Conversely when $r_{\rm nuc}=r_{\rm prop}$, nearly every step produces a nucleation event and climb-over is rare. This implies that $\rho_d \ll 1$ and $\rho_f$ is the probability that the length (modulo $W$) of the graphene segment connected to a B-step is $W-1$ as determined from a Poisson distribution with average value $W \Theta$.

We have solved Eqs.~(\ref{nA})-(\ref{nD}) numerically (assuming $\Delta E'=0$ for simplicity) in the two limits discussed above using the initial conditions
\begin{equation}
n_A = n \hspace{1cm} n_B = n_C = n_D = 0.
\end{equation}
Using this numerical data, we calculate
\begin{equation}
\frac{d\Theta_1}{d\Theta_2} = \frac{d\Theta_1/dt}{d\Theta_2/dt} = \frac{r_{\rm nuc} n_A+ r_{\rm prop} n_B}{r_{\rm nuc} n_C+ r'_{\rm prop} n_D} - 1,  \label{distribution}
\end{equation}
and use $\Theta=\Theta_1+2\Theta_2$ to equate the right side of (\ref{distribution}) to $d\Theta/d\Theta_2-2$.

\begin{figure}
\includegraphics[scale=0.34]{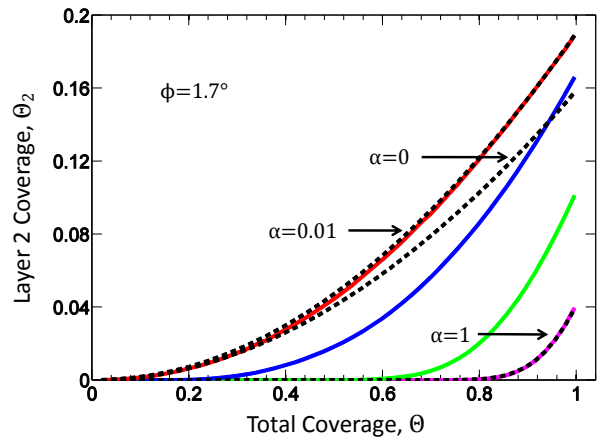}
\caption{The second layer coverage $\Theta_2$ as a function of the total coverage $\Theta$ with $\Delta E'=0$. The solid lines are KMC simulations with (bottom to top) $\Delta E/kT = 0, 3.9, 5.8$ and $11.6$. Dashed Lines are the rate equation results.}
\label{rate}
\end{figure}

Fig.~\ref{rate} compares $\Theta_2$ versus $\Theta$ as determined from the KMC simulation (solid curves) with the correspondingly rate equation results (dashed curves). The agreement is quite good when $\Delta E/kT = 0$ (purple curve). This is the no "climb-over" regime where $\alpha=p_f/(p_f+p_d)=1$. The agreement is similarly good when $\Delta E/kT$ is large (red curve) if we account for coalescence in the rate equations with the choice $\alpha = 0.01$. The no-coalescence curve ($\alpha=0$) falls below the $\alpha \neq 0$ curve because, in the rate equations, the presence of coalescence reduces  the life time for all first layer propagating graphene segments, which  reduce the number of competitors to  second layer propagation. Because  $\Delta E/kT$ is large, there are not many graphene segments in the system to begin with. Removing some first layer segments by coalescence promotes  second layer propagation and thus results in a larger second layer coverage.

\bibliography{Graphene-Growth}

\begin{thebibliography}{17}%
\makeatletter
\providecommand \@ifxundefined [1]{%
 \@ifx{#1\undefined}
}%
\providecommand \@ifnum [1]{%
 \ifnum #1\expandafter \@firstoftwo
 \else \expandafter \@secondoftwo
 \fi
}%
\providecommand \@ifx [1]{%
 \ifx #1\expandafter \@firstoftwo
 \else \expandafter \@secondoftwo
 \fi
}%
\providecommand \natexlab [1]{#1}%
\providecommand \enquote  [1]{``#1''}%
\providecommand \bibnamefont  [1]{#1}%
\providecommand \bibfnamefont [1]{#1}%
\providecommand \citenamefont [1]{#1}%
\providecommand \href@noop [0]{\@secondoftwo}%
\providecommand \href [0]{\begingroup \@sanitize@url \@href}%
\providecommand \@href[1]{\@@startlink{#1}\@@href}%
\providecommand \@@href[1]{\endgroup#1\@@endlink}%
\providecommand \@sanitize@url [0]{\catcode `\\12\catcode `\$12\catcode
  `\&12\catcode `\#12\catcode `\^12\catcode `\_12\catcode `\%12\relax}%
\providecommand \@@startlink[1]{}%
\providecommand \@@endlink[0]{}%
\providecommand \url  [0]{\begingroup\@sanitize@url \@url }%
\providecommand \@url [1]{\endgroup\@href {#1}{\urlprefix }}%
\providecommand \urlprefix  [0]{URL }%
\providecommand \Eprint [0]{\href }%
\providecommand \doibase [0]{http://dx.doi.org/}%
\providecommand \selectlanguage [0]{\@gobble}%
\providecommand \bibinfo  [0]{\@secondoftwo}%
\providecommand \bibfield  [0]{\@secondoftwo}%
\providecommand \translation [1]{[#1]}%
\providecommand \BibitemOpen [0]{}%
\providecommand \bibitemStop [0]{}%
\providecommand \bibitemNoStop [0]{.\EOS\space}%
\providecommand \EOS [0]{\spacefactor3000\relax}%
\providecommand \BibitemShut  [1]{\csname bibitem#1\endcsname}%
\let\auto@bib@innerbib\@empty
\bibitem [{\citenamefont {First}\ \emph {et~al.}(2010)\citenamefont {First},
  \citenamefont {de~Heer}, \citenamefont {Seyller}, \citenamefont {Berger},
  \citenamefont {Stroscio},\ and\ \citenamefont {Moon}}]{First2010}%
  \BibitemOpen
  \bibfield  {author} {\bibinfo {author} {\bibfnamefont {P.~N.}\ \bibnamefont
  {First}}, \bibinfo {author} {\bibfnamefont {W.~A.}\ \bibnamefont {de~Heer}},
  \bibinfo {author} {\bibfnamefont {T.}~\bibnamefont {Seyller}}, \bibinfo
  {author} {\bibfnamefont {C.}~\bibnamefont {Berger}}, \bibinfo {author}
  {\bibfnamefont {J.~A.}\ \bibnamefont {Stroscio}}, \ and\ \bibinfo {author}
  {\bibfnamefont {J.~S.}\ \bibnamefont {Moon}},\ }\href@noop {} {\bibfield
  {journal} {\bibinfo  {journal} {MRS. Bull.}\ }\textbf {\bibinfo {volume}
  {35}},\ \bibinfo {pages} {296} (\bibinfo {year} {2010})}\BibitemShut
  {NoStop}%
\bibitem [{\citenamefont {Sprinkle}\ \emph {et~al.}(2010)\citenamefont
  {Sprinkle}, \citenamefont {Ruan}, \citenamefont {Hu}, \citenamefont
  {Hankinson}, \citenamefont {Rubio-Roy}, \citenamefont {Zhang}, \citenamefont
  {Wu}, \citenamefont {Berger},\ and\ \citenamefont {de~Heer}}]{Sprinkle2010}%
  \BibitemOpen
  \bibfield  {author} {\bibinfo {author} {\bibfnamefont {M.}~\bibnamefont
  {Sprinkle}}, \bibinfo {author} {\bibfnamefont {M.}~\bibnamefont {Ruan}},
  \bibinfo {author} {\bibfnamefont {Y.}~\bibnamefont {Hu}}, \bibinfo {author}
  {\bibfnamefont {J.}~\bibnamefont {Hankinson}}, \bibinfo {author}
  {\bibfnamefont {M.}~\bibnamefont {Rubio-Roy}}, \bibinfo {author}
  {\bibfnamefont {B.}~\bibnamefont {Zhang}}, \bibinfo {author} {\bibfnamefont
  {X.}~\bibnamefont {Wu}}, \bibinfo {author} {\bibfnamefont {C.}~\bibnamefont
  {Berger}}, \ and\ \bibinfo {author} {\bibfnamefont {W.~A.}\ \bibnamefont
  {de~Heer}},\ }\href@noop {} {\bibfield  {journal} {\bibinfo  {journal} {Nat.
  Nanotech.}\ }\textbf {\bibinfo {volume} {5}},\ \bibinfo {pages} {727}
  (\bibinfo {year} {2010})}\BibitemShut {NoStop}%
\bibitem [{\citenamefont {Lauffer}\ \emph {et~al.}(2008)\citenamefont
  {Lauffer}, \citenamefont {Emtsev}, \citenamefont {Graupner}, \citenamefont
  {Seyller}, \citenamefont {Ley}, \citenamefont {Reshanov},\ and\ \citenamefont
  {Weber}}]{Lauffer2008}%
  \BibitemOpen
  \bibfield  {author} {\bibinfo {author} {\bibfnamefont {P.}~\bibnamefont
  {Lauffer}}, \bibinfo {author} {\bibfnamefont {K.~V.}\ \bibnamefont {Emtsev}},
  \bibinfo {author} {\bibfnamefont {R.}~\bibnamefont {Graupner}}, \bibinfo
  {author} {\bibfnamefont {T.}~\bibnamefont {Seyller}}, \bibinfo {author}
  {\bibfnamefont {L.}~\bibnamefont {Ley}}, \bibinfo {author} {\bibfnamefont
  {S.~A.}\ \bibnamefont {Reshanov}}, \ and\ \bibinfo {author} {\bibfnamefont
  {H.~B.}\ \bibnamefont {Weber}},\ }\href@noop {} {\bibfield  {journal}
  {\bibinfo  {journal} {Phys. Rev. B}\ }\textbf {\bibinfo {volume} {77}},\
  \bibinfo {pages} {155426} (\bibinfo {year} {2008})}\BibitemShut {NoStop}%
\bibitem [{\citenamefont {Virojanadara}\ \emph {et~al.}(2008)\citenamefont
  {Virojanadara}, \citenamefont {Syvajarvi}, \citenamefont {Yakimova},
  \citenamefont {Johansson}, \citenamefont {Zakharov},\ and\ \citenamefont
  {Balasubramanian}}]{Virojanadara2008}%
  \BibitemOpen
  \bibfield  {author} {\bibinfo {author} {\bibfnamefont {C.}~\bibnamefont
  {Virojanadara}}, \bibinfo {author} {\bibfnamefont {M.}~\bibnamefont
  {Syvajarvi}}, \bibinfo {author} {\bibfnamefont {R.}~\bibnamefont {Yakimova}},
  \bibinfo {author} {\bibfnamefont {L.~I.}\ \bibnamefont {Johansson}}, \bibinfo
  {author} {\bibfnamefont {A.~A.}\ \bibnamefont {Zakharov}}, \ and\ \bibinfo
  {author} {\bibfnamefont {T.}~\bibnamefont {Balasubramanian}},\ }\href@noop {}
  {\bibfield  {journal} {\bibinfo  {journal} {Phys. Rev. B}\ }\textbf {\bibinfo
  {volume} {78}},\ \bibinfo {pages} {245403} (\bibinfo {year}
  {2008})}\BibitemShut {NoStop}%
\bibitem [{\citenamefont {Hupalo}\ \emph {et~al.}(2009)\citenamefont {Hupalo},
  \citenamefont {Conrad},\ and\ \citenamefont {Tringides}}]{Hupalo2009}%
  \BibitemOpen
  \bibfield  {author} {\bibinfo {author} {\bibfnamefont {M.}~\bibnamefont
  {Hupalo}}, \bibinfo {author} {\bibfnamefont {E.~H.}\ \bibnamefont {Conrad}},
  \ and\ \bibinfo {author} {\bibfnamefont {M.~C.}\ \bibnamefont {Tringides}},\
  }\href@noop {} {\bibfield  {journal} {\bibinfo  {journal} {Phys. Rev. B}\
  }\textbf {\bibinfo {volume} {80}},\ \bibinfo {pages} {041401} (\bibinfo
  {year} {2009})}\BibitemShut {NoStop}%
\bibitem [{\citenamefont {Emtsev}\ \emph {et~al.}(2009)\citenamefont {Emtsev},
  \citenamefont {Bostwick}, \citenamefont {Horn}, \citenamefont {Jobst},
  \citenamefont {Kellogg}, \citenamefont {Ley}, \citenamefont {McChesney},
  \citenamefont {Ohta}, \citenamefont {Reshanov}, \citenamefont {Rohrl},
  \citenamefont {Rotenberg}, \citenamefont {Schmid}, \citenamefont {Waldmann},
  \citenamefont {Weber},\ and\ \citenamefont {Seyller}}]{Emtsev2009}%
  \BibitemOpen
  \bibfield  {author} {\bibinfo {author} {\bibfnamefont {K.~V.}\ \bibnamefont
  {Emtsev}}, \bibinfo {author} {\bibfnamefont {A.}~\bibnamefont {Bostwick}},
  \bibinfo {author} {\bibfnamefont {K.}~\bibnamefont {Horn}}, \bibinfo {author}
  {\bibfnamefont {J.}~\bibnamefont {Jobst}}, \bibinfo {author} {\bibfnamefont
  {G.~L.}\ \bibnamefont {Kellogg}}, \bibinfo {author} {\bibfnamefont
  {L.}~\bibnamefont {Ley}}, \bibinfo {author} {\bibfnamefont {J.~L.}\
  \bibnamefont {McChesney}}, \bibinfo {author} {\bibfnamefont {T.}~\bibnamefont
  {Ohta}}, \bibinfo {author} {\bibfnamefont {S.~A.}\ \bibnamefont {Reshanov}},
  \bibinfo {author} {\bibfnamefont {J.}~\bibnamefont {Rohrl}}, \bibinfo
  {author} {\bibfnamefont {E.}~\bibnamefont {Rotenberg}}, \bibinfo {author}
  {\bibfnamefont {A.~K.}\ \bibnamefont {Schmid}}, \bibinfo {author}
  {\bibfnamefont {D.}~\bibnamefont {Waldmann}}, \bibinfo {author}
  {\bibfnamefont {H.~B.}\ \bibnamefont {Weber}}, \ and\ \bibinfo {author}
  {\bibfnamefont {T.}~\bibnamefont {Seyller}},\ }\href@noop {} {\bibfield
  {journal} {\bibinfo  {journal} {Nat. Mater.}\ }\textbf {\bibinfo {volume}
  {8}},\ \bibinfo {pages} {203} (\bibinfo {year} {2009})}\BibitemShut {NoStop}%
\bibitem [{\citenamefont {Bolen}\ \emph {et~al.}(2009)\citenamefont {Bolen},
  \citenamefont {Harrison}, \citenamefont {Biedermann},\ and\ \citenamefont
  {Capano}}]{Bolen2009}%
  \BibitemOpen
  \bibfield  {author} {\bibinfo {author} {\bibfnamefont {M.~L.}\ \bibnamefont
  {Bolen}}, \bibinfo {author} {\bibfnamefont {S.~E.}\ \bibnamefont {Harrison}},
  \bibinfo {author} {\bibfnamefont {L.~B.}\ \bibnamefont {Biedermann}}, \ and\
  \bibinfo {author} {\bibfnamefont {M.~A.}\ \bibnamefont {Capano}},\
  }\href@noop {} {\bibfield  {journal} {\bibinfo  {journal} {Phys. Rev. B}\
  }\textbf {\bibinfo {volume} {80}},\ \bibinfo {pages} {115433} (\bibinfo
  {year} {2009})}\BibitemShut {NoStop}%
\bibitem [{\citenamefont {Ohta}\ \emph {et~al.}(2010)\citenamefont {Ohta},
  \citenamefont {Bartelt}, \citenamefont {Nie}, \citenamefont {Thurmer},\ and\
  \citenamefont {Kellogg}}]{Ohta2010}%
  \BibitemOpen
  \bibfield  {author} {\bibinfo {author} {\bibfnamefont {T.}~\bibnamefont
  {Ohta}}, \bibinfo {author} {\bibfnamefont {N.~C.}\ \bibnamefont {Bartelt}},
  \bibinfo {author} {\bibfnamefont {S.}~\bibnamefont {Nie}}, \bibinfo {author}
  {\bibfnamefont {K.}~\bibnamefont {Thurmer}}, \ and\ \bibinfo {author}
  {\bibfnamefont {G.~L.}\ \bibnamefont {Kellogg}},\ }\href@noop {} {\bibfield
  {journal} {\bibinfo  {journal} {Phys. Rev. B}\ }\textbf {\bibinfo {volume}
  {81}},\ \bibinfo {pages} {121411} (\bibinfo {year} {2010})}\BibitemShut
  {NoStop}%
\bibitem [{\citenamefont {Joyce}\ \emph {et~al.}(1994)\citenamefont {Joyce},
  \citenamefont {Vvedensky},\ and\ \citenamefont {C.T.}}]{Joyce1994}%
  \BibitemOpen
  \bibfield  {author} {\bibinfo {author} {\bibfnamefont {B.}~\bibnamefont
  {Joyce}}, \bibinfo {author} {\bibfnamefont {D.}~\bibnamefont {Vvedensky}}, \
  and\ \bibinfo {author} {\bibfnamefont {F.}~\bibnamefont {C.T.}},\ }\href@noop
  {} {\emph {\bibinfo {title} {Handbook on Semiconductors: Materials,
  Properties, and Preparation}}},\ edited by\ \bibinfo {editor} {\bibfnamefont
  {S.}~\bibnamefont {Mahajan}},\ Vol.~\bibinfo {volume} {3A}\ (\bibinfo
  {publisher} {Elsevier, Amsterdam},\ \bibinfo {year} {1994})\ pp.\ \bibinfo
  {pages} {275--368}\BibitemShut {NoStop}%
\bibitem [{\citenamefont {Kratzer}\ \emph {et~al.}(2002)\citenamefont
  {Kratzer}, \citenamefont {Penev},\ and\ \citenamefont
  {Scheffler}}]{Kratzer2002}%
  \BibitemOpen
  \bibfield  {author} {\bibinfo {author} {\bibfnamefont {P.}~\bibnamefont
  {Kratzer}}, \bibinfo {author} {\bibfnamefont {E.}~\bibnamefont {Penev}}, \
  and\ \bibinfo {author} {\bibfnamefont {M.}~\bibnamefont {Scheffler}},\
  }\href@noop {} {\bibfield  {journal} {\bibinfo  {journal} {Appl. Phys. A}\
  }\textbf {\bibinfo {volume} {75}},\ \bibinfo {pages} {79} (\bibinfo {year}
  {2002})}\BibitemShut {NoStop}%
\bibitem [{\citenamefont {Hass}\ \emph {et~al.}(2008)\citenamefont {Hass},
  \citenamefont {de~Heer},\ and\ \citenamefont {Conrad}}]{Hass2008}%
  \BibitemOpen
  \bibfield  {author} {\bibinfo {author} {\bibfnamefont {J.}~\bibnamefont
  {Hass}}, \bibinfo {author} {\bibfnamefont {W.~A.}\ \bibnamefont {de~Heer}}, \
  and\ \bibinfo {author} {\bibfnamefont {E.~H.}\ \bibnamefont {Conrad}},\
  }\href@noop {} {\bibfield  {journal} {\bibinfo  {journal} {J. Phys.: Cond.
  Matter}\ }\textbf {\bibinfo {volume} {20}},\ \bibinfo {pages} {323202}
  (\bibinfo {year} {2008})}\BibitemShut {NoStop}%
\bibitem [{\citenamefont {Evans}\ \emph {et~al.}(2006)\citenamefont {Evans},
  \citenamefont {Thiel},\ and\ \citenamefont {Bartelt}}]{Evans2006}%
  \BibitemOpen
  \bibfield  {author} {\bibinfo {author} {\bibfnamefont {J.~W.}\ \bibnamefont
  {Evans}}, \bibinfo {author} {\bibfnamefont {P.~A.}\ \bibnamefont {Thiel}}, \
  and\ \bibinfo {author} {\bibfnamefont {M.~C.}\ \bibnamefont {Bartelt}},\
  }\href@noop {} {\bibfield  {journal} {\bibinfo  {journal} {Surf. Sci. Rept.}\
  }\textbf {\bibinfo {volume} {61}},\ \bibinfo {pages} {1} (\bibinfo {year}
  {2006})}\BibitemShut {NoStop}%
\bibitem [{\citenamefont {Voter}(2007)}]{Voter2007}%
  \BibitemOpen
  \bibfield  {author} {\bibinfo {author} {\bibfnamefont {A.~F.}\ \bibnamefont
  {Voter}},\ }in\ \href@noop {} {\emph {\bibinfo {booktitle} {Radiation Effects
  in Solids}}},\ \bibinfo {series} {NATO Science Series}, Vol.\ \bibinfo
  {volume} {235},\ \bibinfo {editor} {edited by\ \bibinfo {editor}
  {\bibfnamefont {K.~E.}\ \bibnamefont {Sickafus}}, \bibinfo {editor}
  {\bibfnamefont {E.~A.}\ \bibnamefont {Kotomin}}, \ and\ \bibinfo {editor}
  {\bibfnamefont {B.~P.}\ \bibnamefont {Uberuaga}}}\ (\bibinfo  {publisher}
  {Springer Netherlands},\ \bibinfo {address} {Dordrecht},\ \bibinfo {year}
  {2007})\ Chap.~\bibinfo {chapter} {1}, pp.\ \bibinfo {pages}
  {1--23}\BibitemShut {NoStop}%
\bibitem [{\citenamefont {Virojanadara}\ \emph {et~al.}(2010)\citenamefont
  {Virojanadara}, \citenamefont {Yakimova}, \citenamefont {Zakharov},\ and\
  \citenamefont {Johansson}}]{Virojanadara2010}%
  \BibitemOpen
  \bibfield  {author} {\bibinfo {author} {\bibfnamefont {C.}~\bibnamefont
  {Virojanadara}}, \bibinfo {author} {\bibfnamefont {R.}~\bibnamefont
  {Yakimova}}, \bibinfo {author} {\bibfnamefont {A.~A.}\ \bibnamefont
  {Zakharov}}, \ and\ \bibinfo {author} {\bibfnamefont {L.~I.}\ \bibnamefont
  {Johansson}},\ }\href@noop {} {\bibfield  {journal} {\bibinfo  {journal} {J.
  Phys. D: Appl. Phys.}\ }\textbf {\bibinfo {volume} {43}},\ \bibinfo {pages}
  {374010} (\bibinfo {year} {2010})}\BibitemShut {NoStop}%
\bibitem [{Not()}]{Note1}%
  \BibitemOpen
  \href@noop {} {}\bibinfo {note} {The authors thank Prof. Thomas Seyller of
  the University of Erlangen for suggesting this refinement. See also
  Ref.~\protect \rev@citealpnum {Hupalo2009}.}\BibitemShut {Stop}%
\bibitem [{\citenamefont {Michely}\ and\ \citenamefont
  {Krug}(2004)}]{Michely2004}%
  \BibitemOpen
  \bibfield  {author} {\bibinfo {author} {\bibfnamefont {T.}~\bibnamefont
  {Michely}}\ and\ \bibinfo {author} {\bibfnamefont {J.}~\bibnamefont {Krug}},\
  }\href@noop {} {\emph {\bibinfo {title} {Islands, Mounds and Atoms: Patterns
  and Processes in Crystal Growth Far from Equilibrium}}}\ (\bibinfo
  {publisher} {Springer-Verlag, Berlin},\ \bibinfo {year} {2004})\BibitemShut
  {NoStop}%
\bibitem [{\citenamefont {Norimatsu}\ and\ \citenamefont
  {Kusunoki}(2010)}]{Norimatsu2010}%
  \BibitemOpen
  \bibfield  {author} {\bibinfo {author} {\bibfnamefont {W.}~\bibnamefont
  {Norimatsu}}\ and\ \bibinfo {author} {\bibfnamefont {M.}~\bibnamefont
  {Kusunoki}},\ }\href@noop {} {\bibfield  {journal} {\bibinfo  {journal}
  {Physica E}\ }\textbf {\bibinfo {volume} {42}},\ \bibinfo {pages} {691}
  (\bibinfo {year} {2010})}\BibitemShut {NoStop}%
\end{thebibliography}%

\end{document}